\documentstyle[12pt,aasms4]{article}
\def\gtsim{\lower.5ex\hbox{$\; \buildrel > \over \sim \;$}}
\def\ltsim{\lower.5ex\hbox{$\; \buildrel < \over \sim \;$}}

\received{--}
\accepted{--}
\journalid{---}{----}
\articleid{--}{--}

\slugcomment{}

\lefthead{Scharf \& Sahu} \righthead{Lensing and GRB luminosities}

\begin{document}

\title{Lensing and the luminosity of Gamma-ray bursts and their hosts}

\author{C.A. Scharf\altaffilmark{1} and K.C. Sahu\altaffilmark{2} }
\affil{Space Telescope Science
Institute, 3700 San Martin Drive, Baltimore MD 21218, USA}
\altaffiltext{1}{scharf@stsci.edu}
\altaffiltext{2}{ksahu@stsci.edu}

\begin{abstract} 

The quoted intrinsic luminosity of objects has a dependency on the assumed
cosmology, and the (less often considered) assumed gravitational lensing
due to the matter content of the viewing beam; the so called `filled' or
`empty' beam cases. 

We consider the implications of filled vs empty beam assumptions for the
derived total luminosities of recent gamma-ray bursts (GRBs) in which
counterparts at cosmological distances have been confirmed. Conversion
factors between filled and empty beam bolometric luminosities are presented
graphically for a range of cosmologies and redshifts. The tendency for
sources to be most probably demagnified further supports the need for
non-isotropic GRB emission in neutron star merger models, or the need for at
least one massive star in other models. In the most extreme case the true
energy of GRB971214 at $z=3.418$ could be as high as $\sim 8\times 10^{53}$
ergs, a factor $\sim 2$ more than previously estimated.  Similarly, the
effect may account for some of the observed bias towards fainter host
magnitudes.

\end{abstract}


\section{Introduction}

The intrinsic luminosity of an object is estimated using the luminosity
distance to the object. This distance is usually derived assuming a
Robertson-Walker model of the Universe where the matter is uniformly and
homogeneously distributed in space, the so-called filled-beam distance
($d_{L-filled}$).  The matter, as we know it however, is comprised of
compact objects such as stars and galaxies, and has a clearly inhomogeneous
distribution. The opposite extreme to the filled-beam case is the so called
empty-beam case, where all the matter is in the form of compact objects far
from the the line-of-sight;  the derived distance being called the
empty-beam distance ($d_{L-empty}$).  In this latter case the weak lensing
of the rest of the universe de-magnifies the source relative to the
filled-beam case. Although reality may be somewhere between these two
extremes (simulations indeed show that the intermediate case is more
representative of the actual matter distribution, see below) it is clearly
important to see to what extremes the matter distribution can affect the
derived luminosity. 

Recently several authors have re-examined this phenomenon with detailed
investigations of the expected range of magnification effects expected in
various model cosmologies (e.g. ~\cite{wam98}, \cite{kan98}, \cite{hol98}). 
These results, which build on earlier works such as Ostriker \& Vietri
(1986), and Dyer \& Roeder (1973), demonstrate that in most commonly
acceptable cosmologies the probability distribution of the observed
magnitude is strongly peaked intermediate to the empty and filled-beam cases
(the skewness depending largely on whether the lensing matter is distributed
as points or more diffusely, the former skewing magnitudes significantly
towards the empty-beam, de-magnified, case).  However a tail to high
magnification exists, so sources may also be brightened.

In this short paper we point out the relevance of this effect to GRBs and
their host galaxies, we also present a simple diagram for reading off the
extreme case of empty vs filled beam demagnification factors for various
cosmologies. We present the estimated energy output of several GRBs and the
range allowed due to de-magnification effects, as well as the range of
effects on the host galaxy magnitudes. 

\section{Empty vs filled luminosity factors}

Since the bolometric luminosity for a given flux/apparent magnitude varies
as the square of the luminosity distance ($d_L$) we estimate here the
maximum possible difference in inferred luminosity as the ratio
$d_{L-empty}^2/d_{L-filled}^2$. In figures 1 and 2 we plot this luminosity
factor as a function of the cosmological density parameter, $\Omega_0$, and
redshift, $z$, for both open and flat (non-zero cosmological constant,
$\lambda_0\neq 0$) universes. The relevant equations are given in Dyer \&
Roeder (1973) and Fukugita et al (1992).

Several features are worth noting. First, as expected, the higher the
matter density the larger the luminosity factor between empty and filled
beam estimates since the empty beam will be in higher contrast to the
rest of the universe as $\Omega_0$ increases. Second, the luminosity
factors are fairly small for $z\ltsim 1$, at most $\sim 20$\%, but by
$z\sim 4$ can rise as high as $\sim 300$\% (for $\Omega=1$). 

\section{Implications for GRBs}

In Table 1 we summarize the current data on GRBs with identified hosts and
measured redshifts (note: GRB 970228 has a redshift estimated from
photometric colours; Sahu et al. 1997a, Fruchter et al., 1998).  In the
final column we give the range of intrinsic GRB energies if the sources have
been demagnified. These energies represent the extreme case of empty-beam
demagnification for flat ($\Omega_0+\lambda_0=1$) cosmologies with
$\Omega_0=0.2$ or $\Omega_0=1$.  It should be noted that the $\Omega=1$
values are given to illustrate the effect of the overall cosmology.  The
burst GRB971214 is subject to the largest possible underestimate of
intrinsic energy (Odewahn et al. 1998), by a factor of 2, to a maximum of
$\sim 8\times 10^{53}$ ergs.

\section{Host galaxy magnitude distribution}

Recently Hogg \& Fruchter (1998) have shown how the distribution of GRB host
galaxy magnitudes ($24.4\leq R \leq 25.8$, and one with $R=14.3$)  appears
to be biased towards the fainter end of the expected magnitude distribution,
based on measured and inferred redshifts. The bias results in an apparent
deficit in hosts with $21\leq R\ltsim 24$, albeit at low significance. 
While there are clearly other potential reasons for such a bias the
possibility that at least some fraction of these high-z objects have been
de-magnified to lower apparent magnitudes cannot be discounted.  A
luminosity factor of $\sim 1.5$ (a plausible `typical' maximum factor over
the redshift range $1< z <3$, a range expected from the star-formation
history of the Universe (Sahu et al. 1997b; Hogg \& Fruchter, 1998)) results
in a magnitude shift of $\sim -0.44$ which could account for a significant
amount of the faint end bias. Even a more realistic factor of 1.2 ($-0.2$
magnitudes) which might be expected from detailed simulation, e.g. 
Wambsganss, Cen, \& Ostriker (1998) helps alleviate the dearth of brighter
hosts.

\section{Discussion}

Since GRB 971214 is the only GRB with an observed redshift $>2$, this
deserves special mention.  If the luminosity distance for high-z GRBs is
more accurately described by the empty-beam case, then the quoted energy of
GRB971214 could be higher by a factor $\sim 2$ ($\Omega=0.2$), implying an
energy output of some $\sim 8 \times 10^{53}$ ergs.  If the energy is
isotropic, the released energy is too high for most models, including the
neutron star merger models.  This would indicate that the GRB emission may
be strongly beamed, which would be a natural consequence of relativistic
expansion, or may require a neutron star-massive black hole merger model, or
a hypernova model for the GRB. Furthermore, GRB host galaxies to the fainter
side of the expected magnitude distribution could easily be explained by
such a variation in the luminosity distance. 

Detailed studies of cosmological lensing indicate that the most probable
effect is a milder demagnification of high-z sources than these extreme
values, with a probability tail of positive magnification leading up to
strong lensing events. Given the extreme nature of GRBs however, where the
observed energies are already too large compared to the predictions from
many models, the possibility that we are systematically underestimating the
intrinsic energy release is perhaps the more relevant issue. Filled beam
estimates of intrinsic luminosity should therefore be considered subject to
an unknown amount of scatter, tending towards underestimates of the true
value by as much as the factors presented here.

\acknowledgements{M. Donahue is thanked for useful discussions and C.A.S.
acknowledges support through NASA grant NAG5-3257}

\clearpage

\figcaption[]{Contour map of the luminosity factor between empty and filled
beam models for open universes ($\lambda_{0}=0$) as a function of redshift.
The contours are at constant $d_{L-empty}^2/d_{L-filled}^2$ and are labeled.
Dashed contours run from factors 1.1, 1.2, 1.3 to 1.4, solid contours run
from 1.5 in steps of 0.5}

\figcaption[]{Contour map of the luminosity factor between empty and filled
beam models for flat universes ($\Omega_0+\lambda_{0}=1$) as a function of
redshift. Contours are as for Figure 1}

\clearpage

\begin{table*}
\begin{center}
\begin{tabular}{lcccc}
Name  & Redshift & Energy (ergs)\tablenotemark{a}& 
Maximum energy `empty-beam' \\
   &   & $(\Omega_{0}=1)$, $(\Omega_0=0.2$, $\lambda_0=0.8)$ &  $(\Omega_{0}=1)$,
$(\Omega_0=0.2$, $\lambda_0=0.8)$ \\   
\tableline
980703\tablenotemark{b}   &   0.966   & $ 0.7 - 1.4\times10^{53}$  & $0.9 - 1.5
\times10^{53}$\\
971214\tablenotemark{c}   &   3.418   & $ 1.44 - 4\times10^{53}$  &$3.8   -  8
\times10^{53} $\\ 
970508\tablenotemark{d} & 0.835 & $  5.18 - 9.1\times 10^{51}$ &$6.22 - 10
\times10^{51}$\\
970228\tablenotemark{e}   &   $\ge 1$ & $ \geq 0.69 - 1.4 \times10^{51} $ &$0.9
-1.54\times10^{51}$\\ 
980425\tablenotemark{f} & 0.0085 & $ 8.1\times 10^{47}$ & $ 8.1\times
10^{47}$ \\
\end{tabular}
\end{center}
\tablenotetext{a}{Derived energies assuming $H_0=65$ kms$^{-1}$ } 
\tablenotetext{b}{\cite{blo98a}}
\tablenotetext{c}{\cite{kul98}}
\tablenotetext{d}{\cite{met97}}
\tablenotetext{e}{\cite{par97}}
\tablenotetext{f}{\cite{blo98b}}
\tablenum{1}
\tablecomments{GRBs with identified hosts and measured redshifts. Full
and empty beam energies are estimated for $\Omega_0=1$ and $\Omega_0=0.2$,
$\lambda_0=0.8$ cosmologies.} \end{table*}

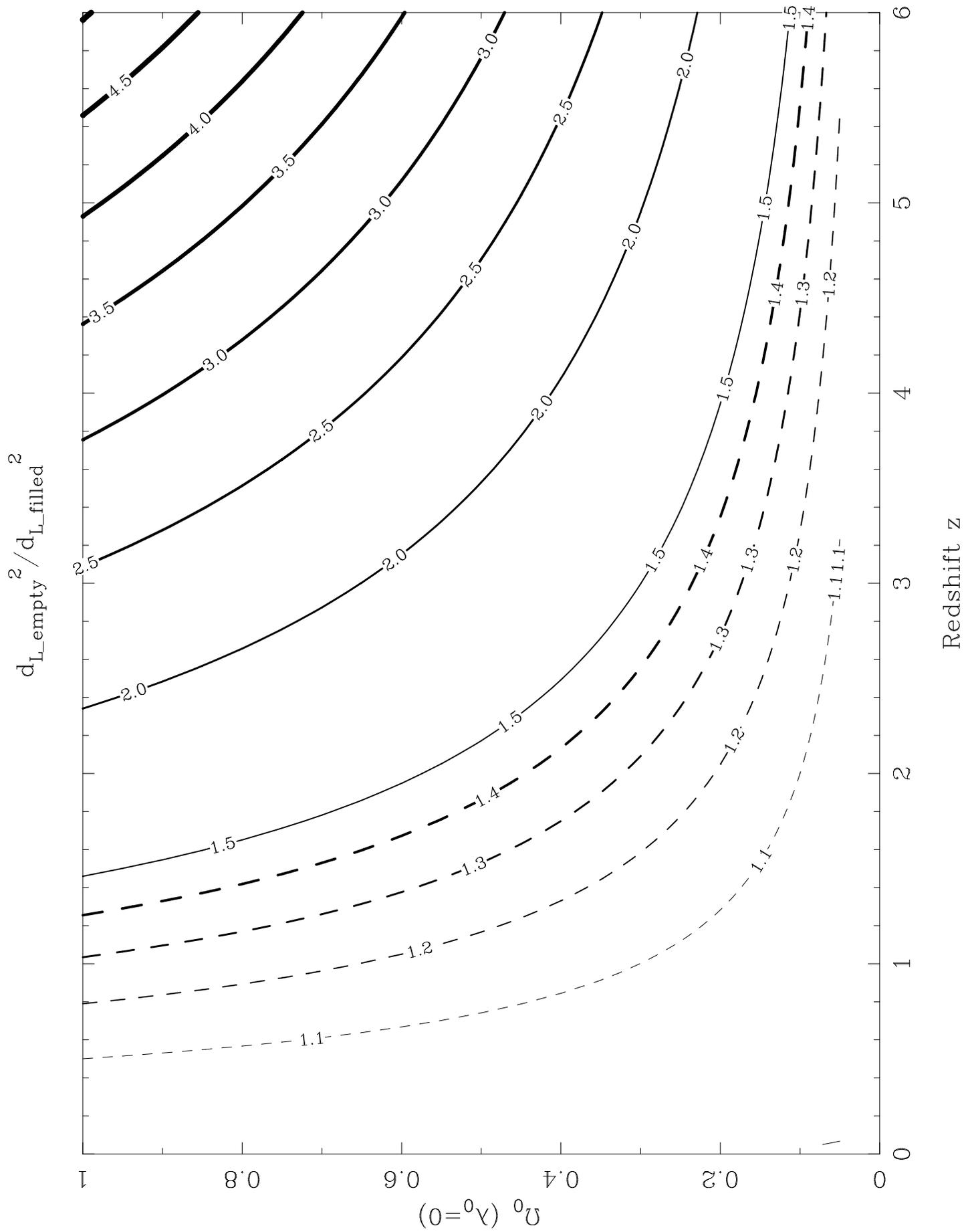
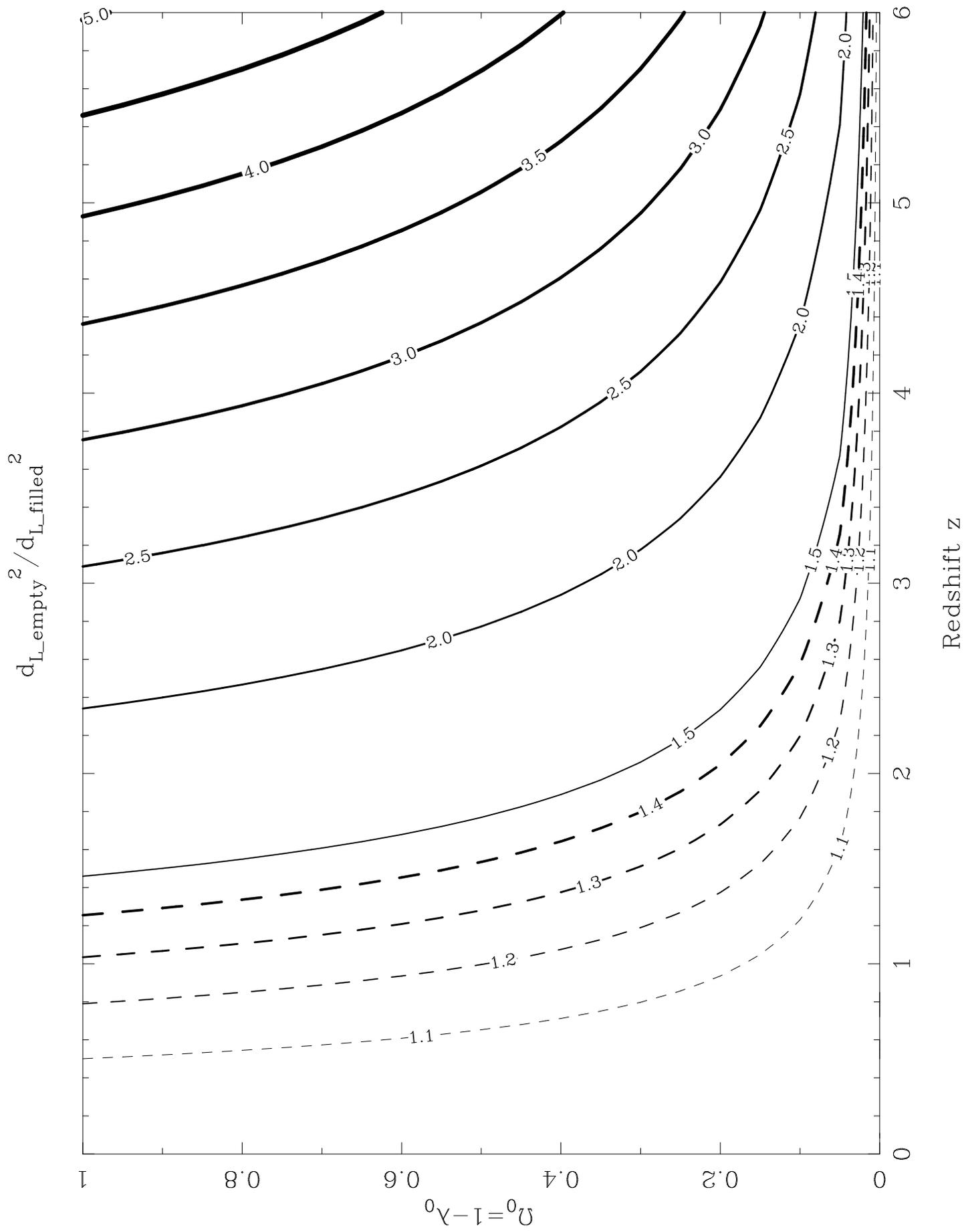
\end{document}